\documentclass[
 reprint,
amsmath,amssymb,aps
,pre
]{revtex4-1}

\usepackage{graphicx}
\usepackage{lineno}
\usepackage{dcolumn}
\usepackage{bm}
\usepackage[normalem]{ulem}
\usepackage{xcolor}

\begin{document}

\title{The frustrated Ising model on the body-centered cubic lattice}

\author{M. Schmidt}
\email{mateus.schmidt@ufsm.br}
\affiliation{Departamento de F\'isica, Universidade Federal de Santa Maria, 97105-900 Santa Maria, RS, Brazil}
\author{G. L. Kohlrausch}
\affiliation{Instituto de F\'isica, Universidade Federal do Rio Grande do Sul, 91501-970 Porto Alegre, RS, Brazil}
\author{F. M. Zimmer}
\affiliation{%
Instituto de F\'isica, Universidade Federal de Mato Grosso do Sul, 79070-900 Campo Grande, MS, Brazil
}%

\begin{abstract}
Recent results for the  Ising model with first ($J_1$) and second ($J_2$) neighbour interactions on the body-centered cubic (bcc) lattice suggest that this model can host signatures of strong frustration, including Schottky anomalies and residual entropy, as well as, a spin-liquid-like phase  [E. Jur\v{c}i\v{s}inov\'a and M. Jur\v{c}i\v{s}in, Phys. Rev. B, {\bf 101}  214443 (2020)].
Motivated by these findings, we investigate phase transitions and thermodynamics of this model using a cluster mean-field approach.
In this lattice, tuning $g=J_2/J_1$ leads to a ground-state transition between antiferromagnetic (AF) and superantiferromagnetic (SAF) phases at the frustration maximum $g=2/3$. Although the ordering temperature is reduced as $g \to 2/3$, our findings suggest the absence of any Schottky anomaly and residual entropy, in good agreement with Monte Carlo simulations. We also find a direct transition between AF and SAF phases, ruling out the presence of the spin-liquid-like state. 
Furthermore, the cluster mean-field outcomes support a scenario with only continuous phase transitions between the paramagnetic state and the low-temperature long-range orders. 
Therefore, our results indicate the absence of strong frustration effects in the thermodynamics and in the nature of phase transitions, which can be ascribed to the higher dimensionality of the bcc lattice.

\end{abstract} 
\maketitle

\section{Introduction}

Spin models with competing exchange couplings between first ($J_1$) and second ($J_2$) neighbours are a permanent venue to investigate frustration effects on phase transitions. 
The complex phenomena hosted by $J_1$-$J_2$ spin models has motivated numerous investigations 
in the past few years \cite{PhysRevB.101.214443,MURTAZAEV2019669, PhysRevE.98.022123, SOROKIN20183455, PhysRevE.99.012134, SCHMIDT2021168151, ZUKOVIC2021127405}. Recently, the thermal phase transitions presented by the antiferromagnetic (AF) $J_1$-$J_2$ Ising model on the body-centered cubic (bcc) lattice has been a relevant subject of debate \cite{PhysRevB.101.214443, MURTAZAEV2019669}. When both interactions  are AF, the ground-state can change from the Neel antiferromagnetic state (for $0<g<2/3$) to a superantiferromagnetic (SAF) state ($g > 2/3$), by modifying the ratio $g \equiv J_2/J_1$. Central issues concern the phase transitions near the maximum frustration point ($g=2/3$). In particular, relevant questions are whether the model host tricriticality at the transition between SAF and paramagnetic (PM) phases, as well as, the possible onset of a spin-liquid-like phase \cite{PhysRevB.101.214443, MURTAZAEV2019669}. Motivated by the rich phenomenology recently reported for the $J_1$-$J_2$ Ising model on the bcc lattice, in the present work, we investigate the thermal phase transitions and the thermodynamics hosted by this model.

In a recent work, the phase diagram of the AF $J_1$-$J_2$ Ising model on the bcc lattice was investigated within Monte Carlo simulations \cite{MURTAZAEV2019669}. By employing the energy histogram analysis, the authors found indication of first-order phase transitions for $2/3 < g \leq 0.75$ and second-order phase transitions for $0 \leq g \leq 2/3$ and $0.8 \leq g \leq 1$. The presence of first-order phase transitions in the PM-SAF phase boundary of the model were also suggested by previous Monte Carlo studies \cite{PhysRevB.20.3820, SOROKIN20183455}, although the precise location of a possible tricritical point remains elusive. 

On the other hand, recent calculations within a recursive lattice approach suggest only second-order phase transitions at the PM-SAF phase boundary \cite{PhysRevB.101.214443}, in agreement with the standard mean-field calculations for the model \cite{PhysRevB.20.3820, Katsura_1974}. Signatures of a spin-liquid-like state between the low-temperature AF and SAF phases were also reported within the recursive lattice approach. In addition, the authors found signatures of strong frustration in the thermodynamic quantities of the model. For instance, round maxima in the specific heat are found within both ordered and disordered phases. Moreover, a finite zero-temperature entropy for the intermediate and SAF states indicate that the model can host a macroscopically degenerate ground-state. It is worth to note, that these features are absent in previous reports for the thermodynamics of the model \cite{murtazaev2018studying}. Therefore, further investigations are required to address both the nature of phase transitions and the thermodynamics of the model, specially near $g=2/3$.

\begin{figure*}[t]
\centering\includegraphics[width=0.9\textwidth]{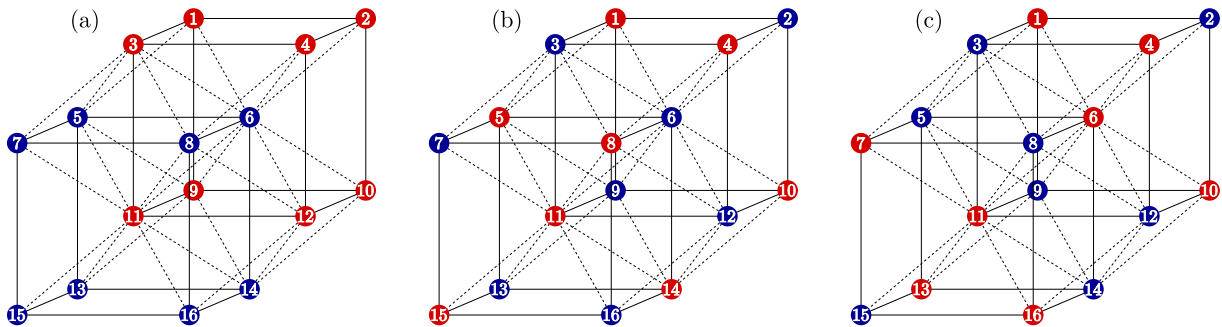}
\caption{Cluster adopted in the CMF calculations. First- and second-neighbour interactions are denoted by dashed and solid lines, respectively. For simplicity, the mean fields are omitted. Colors of circles indicate the pattern of the (a) AF and (b)-(c) SAF phases. }\label{fig:cmf}
\end{figure*}

In this paper, we investigate the AF $J_1$-$J_2$ Ising model on the bcc lattice within a cluster mean-field (CMF) approach. This method has been used in several recent studies of phase diagrams and thermodynamic quantities of frustrated spin models \cite{PhysRevB.87.144406, GUERRERO2020167140, PhysRevE.99.012134, Schmidt_2015, Alavani_2018, PhysRevB.91.174424, PhysRevE.102.032139, PhysRevB.96.014431, PhysRevLett.125.057204, PhysRevB.100.140410,Ren_2014,PhysRevLett.112.127203, PhysRevLett.114.027201,FGODOY2020126687,SCHMIDT2021168151,PhysRevE.103.032125}. In particular, this framework provided a description of phase transitions for $J_1$-$J_2$ spin models on square \cite{PhysRevB.87.144406, FGODOY2020126687} and honeycomb \cite{SCHMIDT2021168151} lattices in very good agreement with recent Monte Carlo results. In addition, mean-field-like methods are expected to provide better results as dimensionality increases, which indicates that the CMF approach can provide a proper description for the $J_1$-$J_2$ bcc lattice. Moreover, the CMF method allows us to thoroughly investigate the thermodynamics of the model, which is a worth subject given the recent findings for entropy and specific heat of the model \cite{PhysRevB.101.214443}.

The rest of the paper is organized as follows. In Sec. \ref{sec2}, we present the model and the CMF theory. The results, including the coupling-temperature phase diagram and thermodynamic quantities, are presented and discussed in Sec. \ref{sec3}. We present
our conclusion in Sec. \ref{sec4}.

\section{Model}\label{sec2}

In this work, we investigate the following Hamiltonian
\begin{equation}
    H = J_1 \sum_{\langle i,j \rangle}\sigma_i \sigma_j +J_2 \sum_{\langle\langle i,j \rangle\rangle} \sigma_i \sigma_j,
\end{equation}
where $\sigma_i=\pm 1$ are Ising spins at the vertices of a bcc lattice and $\langle{i,j} \rangle$ and $\langle\langle{i,j} \rangle\rangle$ denotes a sum over pairs of first- and second-neighbours, respectively. We also adopt antiferromagnetic exchange couplings and, therefore, $J_1>0$ and $J_2\geq0$. For weak $J_2$, the model exhibits an antiferromagnetic ground-state (see Fig. \ref{fig:cmf}(a)), with energy per particle given by $u_{AF}=-4J_1+3J_2$. For stronger $J_2$, the model can exhibit a SAF ground-state, which is four-fold degenerated. The two SAF ground states that cannot be related by global spin inversion are illustrated in Fig. \ref{fig:cmf}(b)-(c), which exhibit the same ground-state energy per particle $u_{SAF}=-3J_2$. At $g=2/3$, a ground-state transition between AF and SAF phases occurs.

In order to investigate the effects of thermal fluctuations on this model, we employ a cluster mean-field approach. This method has been widely employed in the study of several spin systems and can also be derived within a variational approach \cite{PhysRevB.87.144406}. Within the CMF framework, the system is divided into identical finite clusters, where the interactions inside a given cluster are incorporated exactly and the intercluster interactions are evaluated by following the usual mean-field approximation
\begin{equation}
    \sigma_i \sigma_j \approx \sigma_i \langle\sigma_j\rangle + \langle \sigma_i \rangle \sigma_j - \langle\sigma_i\rangle \langle\sigma_j\rangle.
\end{equation}
In the present work, we propose a cluster of size $n_s=16$ shown in Fig. \ref{fig:cmf}, which allows us to investigate both AF and SAF phases and also contains the same number of sites on each sublattice formed within both AF and SAF phases.  Therefore, the CMF Hamiltonian for a single-cluster $\nu$ is given by
\begin{equation}\begin{split}
    H_{\textrm{CMF}}= J_1 \sum_{\langle i,j \in \nu \rangle}\sigma_i \sigma_j +J_2 \sum_{\langle\langle i,j \in \nu \rangle\rangle} \sigma_i \sigma_j \\ 
    + J_1 \sum_{\langle i,j \rangle'}\left(\sigma_i m_j -\frac{m_i m_j}{2}\right) 
    \\ +J_2 \sum_{\langle\langle i,j \rangle\rangle'} \left(\sigma_i m_j -\frac{m_i m_j}{2} \right), \label{eq:ham_CMF}
    \end{split}
\end{equation}
where $\langle i,j \rangle'$ and $\langle \langle i,j \rangle \rangle'$ denote sums over sites $i$, that belong to the cluster $\nu$, and $j$, that belongs to neighbour clusters. In addition, 
\begin{equation}
m_i = \langle \sigma_i \rangle = \frac{\textrm{Tr} \, \sigma_i \, e^{- H_{\textrm{CMF}}/k_B T}}{\textrm{Tr} \,  e^{- H_{\textrm{CMF}}/k_BT}} \label{eq:local_mag}    \end{equation}
is the local magnetization of the site $i$, where $k_B$ is the Boltzmann constant and $T$ is temperature. Considering the pattern of AF and SAF phases, the behavior of local magnetizations from equivalent sites on different clusters should be identical. Therefore, one can evaluate the local magnetizations coming from the neighbourhood of $\nu$ by computing it from equivalent sites within $\nu$.

After solving equations (\ref{eq:ham_CMF}) and (\ref{eq:local_mag}) self-consistently, one can compute the thermodynamics of the model. For instance, the free energy per spin,
\begin{equation}
    f = -\frac{k_B T}{n_s} \ln\left(\textrm{Tr} \,e^{- H_{\textrm{CMF}}/k_B T} \right),
\end{equation}
 and the internal energy per spin,
\begin{equation}
u = \frac{1}{n_s} \frac{\textrm{Tr} \, H_{\textrm{CMF}} \, e^{- H_{\textrm{CMF}}/k_B T}}{\textrm{Tr} \,  e^{- H_{\textrm{CMF}}/k_BT}},
\end{equation}
can be evaluated in a straightforward way. From this quantities, one can compute the entropy per spin
\begin{equation}
s=(u-f)/T.    
\end{equation}
and also investigate the specific heat
\begin{equation}
c_v = \frac{d u}{d T}.
\end{equation}

The AF phase can be described by a non-zero staggered magnetization following the pattern indicated in Fig. \ref{fig:cmf}(a). The behavior of local magnetizations allow us to identify the following order parameter for the AF phase: 
$m_{AF} = (m_1+m_2+m_3+m_4-m_5-m_6-m_7-m_8+m_{9}+m_{10}+m_{11}+m_{12}-m_{13}-m_{14}-m_{15}-m_{16})/16$. The SAF phase can be found in one of the two states depicted in panels (b) and (c) of Fig. \ref{fig:cmf}. Within our CMF calculations, we found that the free energy of the state described in panel (c) is lower than that of the state shown in panel (b) at finite temperatures. This distinction could be attributed to the difference in the number of frustrated first-neighbour interactions incorporated within the cluster for each state. It is worth to note that this difference is solely a consequence of the CMF approximation, instead of a feature of the model. In the following, we build the phase diagrams by considering only the state described in panel (c), as it shows a lower free-energy. Then, the SAF state can be described by the following order parameter: $m_{SAF} = (m_1-m_2-m_3+m_4-m_5+m_6+m_7-m_8-m_{9}+m_{10}+ m_{11}-m_{12}+m_{13}-m_{14}-m_{15}+m_{16})/16$. 

In Ref. \cite{PhysRevB.101.214443}, an intermediate state is found between AF and SAF phases. The authors differ this state from the AF and SAF phases by adopting a three-sublattice structure. By comparing the sublattice magnetizations, the transition to the intermediate state can be identified, which is also signaled by a discontinuity in the specific heat. Within our CMF calculations, the possibility of such an intermediate state can be tested by comparing local magnetizations from topologically equivalent sites. For instance, the intermediate state can be present whether $|m_6| \neq |m_{11}|$, which is not expected for the AF phase, in which $m_6=-m_{11}$, neither in the SAF phase, in which $m_6=m_{11}$. In order to allow a possible onset of the intermediate state, we compute the 16 local magnetizations as independent parameters. More importantly, the behavior of the specific heat also should signal the onset of the intermediate state.

\section{Results and discussion}\label{sec3}

\begin{figure}[t]
\centering\includegraphics[width=0.99\columnwidth]{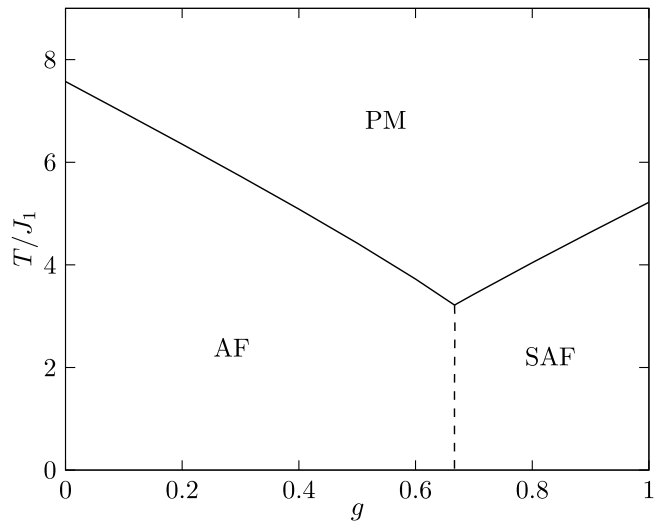}
\caption{Temperature-coupling phase diagram, where the solid lines indicate second-order phase transitions and the dashed one denotes first-order phase transitions.}\label{fig:pd}
\end{figure}

In the following, we discuss our numerical findings, in which, for simplicity, we adopted $k_B=1$. The AF phase is characterized by $m_{AF}\neq0$ and $m_{SAF} = 0$ while the SAF phase is identified when  $m_{AF}=0$ and $m_{SAF} \neq 0$, with the PM state occurring for
 $m_{AF}=m_{SAF}=0$. In Fig. \ref{fig:pd}, our results for the coupling {\it versus} temperature phase diagram of the model are shown. Our CMF calculations reproduce correctly the ground-state transition between AF and SAF phases at $g=2/3$. Moreover, the temperature in which order-disorder phase transitions take place are reduced as $g=2/3$ is approached. This reduction in the ordering temperature is a consequence of the frustration introduced by the competition between first- and second-neighbour couplings. However, it is worth to note that the order-disorder phase transitions at $g=2/3$ still take place at finite temperature, in agreement with recent findings for the model \cite{MURTAZAEV2019669,PhysRevB.101.214443}. It means that the effects of frustration on the bcc lattice are weaker when compared to those found for the frustrated square and honeycomb lattices, in which antiferromagnetic second-neighbour interactions can bring the ordering temperature to zero \cite{PhysRevB.86.134410, kalz2008phase, PhysRevE.91.032145, SCHMIDT2021168151, ZUKOVIC2021127405, BOBAK20162693}. A possible explanation for this reduced sensitivity to frustration can be related to the higher dimensionality of the bcc lattice, which leads to long-range orders that are more robust under thermal fluctuations even when second-neighbour couplings introduce frustration. Analogous findings for the frustrated square lattice with interlayer couplings also support this hypothesis \cite{FGODOY2020126687, murtazaev2017critical}.  
 
In addition, our findings support that only second-order phase transitions take place at the phase boundaries between the PM state and the low temperature phases. For the PM-AF phase boundary, the continuous nature of the phase transitions is also supported by several studies \cite{PhysRevB.20.3820, PhysRevB.27.401, MURTAZAEV2019669,PhysRevB.101.214443}. In particular, our estimate for the reduced critical temperature in the limit case $g=0$ is $T_N/J_1=7.573$. This result represents an improvement over the single-site mean field approach, that delivers $T_N/J_1=8$, when compared to large scale Monte Carlo calculations \cite{MC_bcc}, in which $T_N/J_1=6.354$ is found. Moreover, our CMF result is much closer to the Monte Carlo one when compared to the finding from Ref. \cite{PhysRevB.101.214443}, where the analysis of the phase diagram allows us to roughly estimate $T_N/J_1 \approx 2.5$ for this limit case.
It is worth noting that Monte Carlo simulations \cite{MURTAZAEV2019669} indicate the presence of both first- and second-order phase transitions for the PM-SAF phase boundary, while recent analytical results \cite{PhysRevB.101.214443} indicate the absence of discontinuous phase transitions in this phase boundary. Therefore, our findings support the second scenario, in which only second-order PM-SAF phase transitions can be found in the $J_1$-$J_2$ Ising model on the bcc lattice. 
\begin{figure}[t]
\centering\includegraphics[width=0.99\columnwidth]{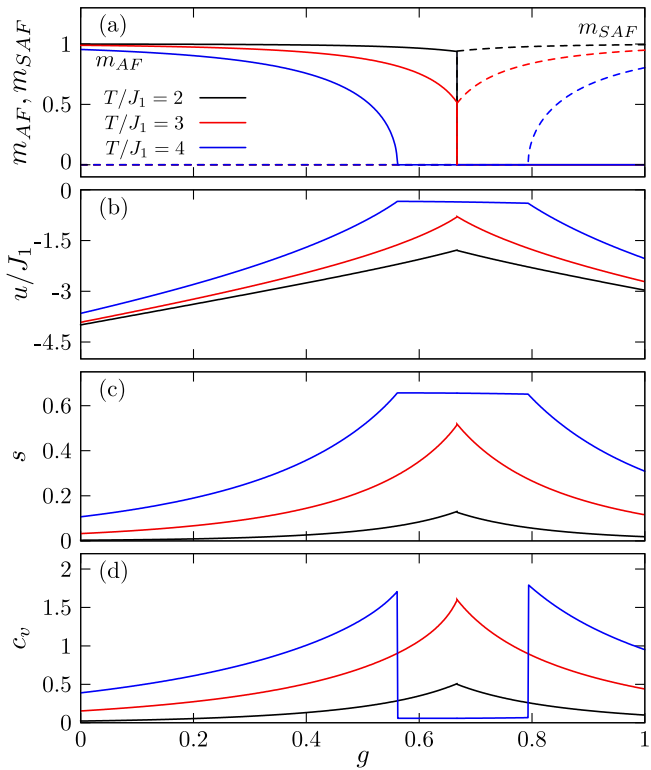}
\caption{Behavior of (a) order parameters, (b) internal energy per spin, (c) entropy per spin and specific heat as function of the coupling ratio for several temperatures. The SAF order parameter is shown with dashed lines in panel (a).}\label{fig:thermo_vs_g}
\end{figure}

By comparing free-energies of different solutions, we were also able to locate the AF-SAF phase boundary, in which we found only first-order phase transitions. At this phase boundary, order parameters, entropy and internal energy can show discontinuities, as shown in Fig. \ref{fig:thermo_vs_g} for $T/J_1=2$ and $3$. We remark that the discontinuities observed for entropy and internal energy are less noticeable when compared to the ones found for both order parameters $m_{AF}$ and $m_{SAF}$.  These small discontinuities in $s(g)$ and $u(g)$ are expected given that the AF-SAF phase boundary is almost a straight vertical line in the phase diagram. In fact, if the slope of $g(T)$ at the transition is zero, the entropy of both phases under transformation is expected to be the same \cite{stillinger2001kauzmann, PhysRevLett.95.087201}.

\begin{figure}[t]
\centering\includegraphics[width=0.9\columnwidth]{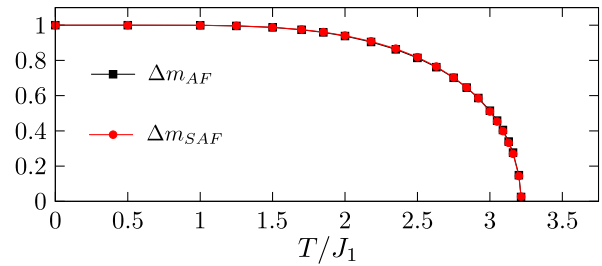}
\caption{Jumps of the order parameters at the AF-SAF first-order phase transition as a function of temperature.}\label{fig:jump}
\end{figure}

\begin{figure}[b]
\centering\includegraphics[width=0.99\columnwidth]{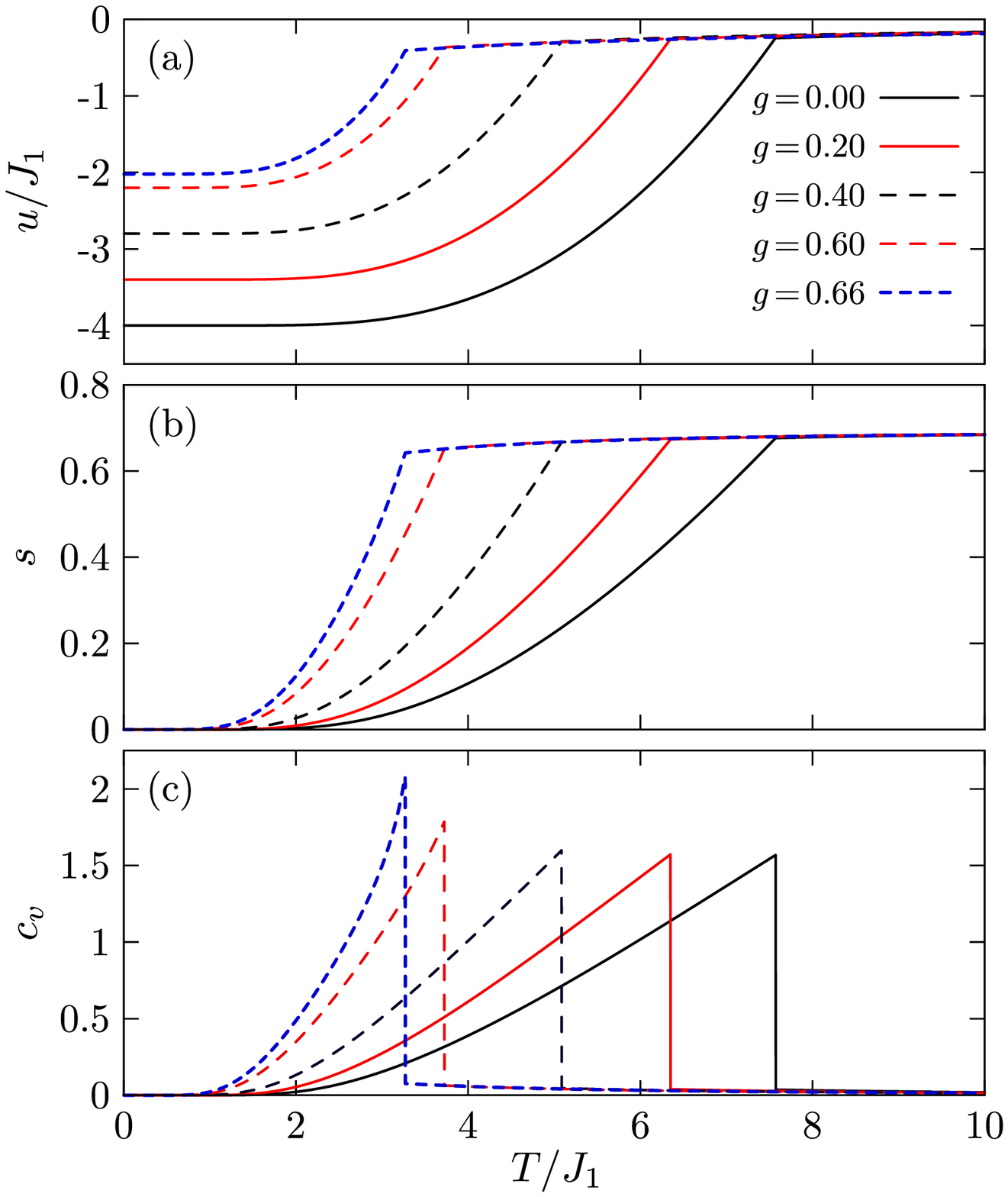}
\caption{Thermal dependence of the (a) internal energy, (b) entropy and (c) specific heat near the PM-AF phase transition for several values of the coupling ratio $g$.}\label{fig:thermo1}
\end{figure}

\begin{figure}[t]
\centering\includegraphics[width=0.97\columnwidth]{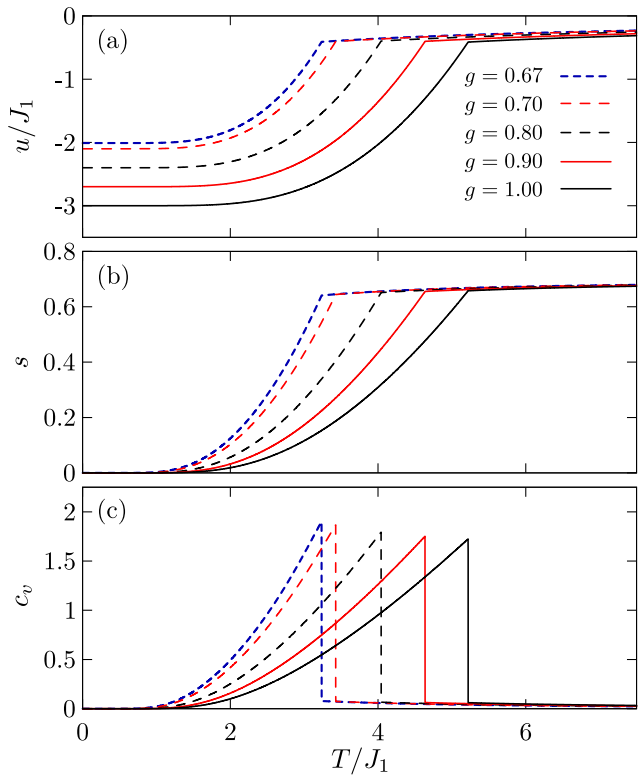}
\caption{Thermal dependence of the (a) internal energy, (b) entropy and (c) specific heat near the PM-SAF phase transition for several values of the coupling ratio $g$.}\label{fig:thermo2}
\end{figure}

The AF-SAF phase boundary meets the PM-AF and PM-SAF phase boundaries at coordinates $T^*/J_1 \approx 3.215$ and $g^* \approx 0.667$ in the coupling-temperature plane. Therefore, a change in the nature of phase transitions occur at this point. One can spot this change in the nature of phase transitions by evaluating the discontinuous jumps of $m_{AF}$ and $m_{SAF}$ at the phase boundary, which are more noticeable when compared to the jumps in other quantities, as discussed above. In Fig. \ref{fig:jump}, we present the jumps of the order parameters at the AF-SAF phase transition for different temperatures. 
As a result, we find that the jumps of both AF and SAF order parameters decrease with temperature, approaching zero as $T \to T^*$ from low temperatures.  For $T>T^*$, only continuous order-disorder phase transitions are observed and, therefore, an absence of discontinuities in the order parameters is expected (see, for instance, the results for $T/J_1=4$ in Fig. \ref{fig:thermo_vs_g} (a)). Therefore, the results shown in Fig. \ref{fig:jump} are consistent with the change in the nature of phase transitions found for the model at $T^*$.

To investigate a possible second-order phase transition to an intermediate state, as pointed in Ref. \cite{PhysRevB.101.214443}, we can search for a signal in the thermodynamics of the model.
For instance, the transitions between the PM state and  the ordered states (AF and SAF) is indicated by an abrupt change in the behavior of the internal energy and entropy, which can drive a discontinuity in specific heat, as shown for $T/J_1=4$ in Fig. \ref{fig:thermo_vs_g}. However, no signature of a transition between the ordered states (AF and SAF) and a possible intermediate state is found within our calculations for the thermodynamic quantities, as shown for $T/J_1=2$ and $3$ in Fig. \ref{fig:thermo_vs_g}. Furthermore, by comparing the local magnetizations from topologically equivalent sites within the cluster, no indication of an intermediate state was noticed. Therefore, our CMF results suggest that the proposed spin-liquid-like state is not realized in this model, which is in agreement with recent Monte Carlo results \cite{murtazaev2018studying}.

In order to provide benchmarks for the present model, we compute the thermal dependence of relevant thermodynamic quantities for several values of $g$. In Figs. \ref{fig:thermo1} and \ref{fig:thermo2}, we present our findings for the internal energy per spin, entropy per spin and specific heat near the PM-AF and PM-SAF phase transitions, respectively. We note that the thermodynamics is characteristic of second-order phase transitions, with the continuous behavior of internal energy and entropy, as well as, the discontinuity in the specific heat. As a result, we find that the increase in $g$ leads to an increase in the internal energy within the AF state. In particular, the ground-state internal energy per spin agrees with the description proposed in Sec. \ref{sec2}. 

We remark that the data presented in Ref. \cite{PhysRevB.101.214443} suggests that the SAF phase exhibits a finite ground-state entropy, which is often related to high levels of frustration. It is worth to note that the SAF ground-state is four-fold degenerate, but no macroscopic degeneracy is expected \cite{MURTAZAEV2019669, murtazaev2018studying}.  As shown in Fig \ref{fig:thermo2} (b), the entropy vanishes at zero temperature, confirming the expected absence of residual entropy in the SAF ground-state. Therefore, the residual entropy found in Ref. \cite{PhysRevB.101.214443} can be pointed as an artifact of the theoretical framework adopted, instead of a feature of the model.
In addition, our results indicate the absence of the Schottky-type anomalies in the specific heat found in Ref. \cite{PhysRevB.101.214443}. It is worth to note that the presence of a round maximum on the specific heat is often pointed as a signature of strong frustration and is found in several frustrated spin systems \cite{canals2016fragmentation, PhysRevE.75.061118, SEMJAN2020126615}. The absence of these anomalies suggests that the competing interactions are unable to introduce strong frustration effects on the $J_1$-$J_2$ Ising model on the bcc lattice. Furthermore, the presence of a single discontinuity on the specific heat for $g=0.66$ and $0.67$ reinforce the absence of the intermediate state near $g=2/3$, which is proposed in the same study.

\section{Conclusion} \label{sec4}

We present a CMF study of the AF $J_1$-$J_2$ Ising model on the bcc lattice. We find only second-order phase transitions at both PM-AF and PM-SAF phase boundaries and first-order phase transitions between AF and SAF phases. We also note that thermodynamic quantities lack any signature of the spin-liquid-like state proposed in Ref. \cite{PhysRevB.101.214443}. Moreover, our data indicate the absence of signatures of strong frustration, such as a round maximum in the specific heat and residual entropy, in agreement with Monte Carlo results \cite{murtazaev2018studying}. 

Our findings also suggest the absence of tricriticality, which is in agreement with several numerical and analytical results for the PM-AF phase boundary, although the nature of the PM-SAF phase transitions still remains elusive. In particular, Monte Carlo simulations \cite{MURTAZAEV2019669} indicate that a tricritical point can be found in the PM-SAF phase boundary, while the results from Ref. \cite{PhysRevB.101.214443} corroborate our findings. Therefore, further investigations are required to clarify the presence of a tricritical point in the PM-SAF phase boundary. 

In summary, our study indicates that the competitive interactions on the $J_1$-$J_2$ bcc lattice are unable to drive strong frustration effects in the thermodynamics and order-disorder phase transitions hosted by the model. This picture differs from the one reported for the same model on two-dimensional systems, such as the square \cite{PhysRevB.87.144406} and honeycomb lattices \cite{SCHMIDT2021168151,ZUKOVIC2021127405}, in which stronger frustration effects take place. In our opinion, the higher dimensionality of the bcc lattice can be responsible for weakening frustration effects in the present model.

\section*{Acknowledgments}
This work was supported by the brazilian agency Conselho Nacional de Desenvolvimento Cient\'ifico e Tecnol\'ogico (CNPq), Coordena\c{c}\~ao de Aperfei\c{c}oamento de Pessoal de N\'ivel Superior (Capes) and Funda\c{c}\~ao de Amparo \`a Pesquisa do Estado do Rio Grande do Sul (Fapergs).

\bibliography{referencia.bib}

\end{document}